\documentclass[12pt]{article}
\newcommand{\text}{\rm}
\begin{document}

\title{\textbf{Vortex Correlation Functions in Maxwell-Chern-Simons Models}}
\author{L. E. Oxman, S.P. Sorella \and $\;$\vspace{4mm}\textbf{\ } \\
%EndAName
{\small {\textit{UERJ, Universidade do Estado do Rio de Janeiro}}} \\
{\small {\textit{Instituto de F\'{i}sica, Departamento de F\'{i}sica
Te\'{o}rica, }}} \\
{\small {\textit{\ Rua S\~{a}o Francisco Xavier 524, 20550-013 Maracan\~{a},
Rio de Janeiro, Brazil}}}\vspace{2mm}}
\maketitle

\begin{abstract}
Maxwell-Chern-Simons models in the presence of an instanton anti-instanton
background are studied. The saddle-point configuration corresponds to the
creation and annihilation of a vortex localized around the Dirac string
needed to support the nontrivial background.

This configuration is generalized to the case in which a nonlocal Maxwell
term is allowed in order to fulfill the finite action requirement.

Following 't Hooft procedure, we compute the vortex correlation functions
and we study the possibility of obtaining spin $1/2$ excitations. A possible
connection with the bosonization of interacting three-dimensional massive
fermionic systems is also discussed.
\end{abstract}

%\LARGE{

\vfill\newpage\ \makeatother

\renewcommand{\theequation}{\thesection.\arabic{equation}}

\section{Introduction}

Bosonization is an important tool to study interacting fermionic systems.
Concerning the case of parity breaking models in $(2+1)D$, many efforts are
being undertaken in order to improve this program. In particular, it is well
established that the correlation functions of $U(1)$ fermionic currents
correspond to correlation functions of topological currents in the dual
bosonized theory \cite{b,b11}. This feature holds for both $(1+1)D$ and $%
(2+1)D$ models and has a universal character \cite{b1}, as stated by the
following formula 
\begin{equation}
K_{F}[\psi ]+I[j^{F}]\leftrightarrow K_{B}[\lambda ]+I[\varepsilon \partial
\lambda ]  \label{univ}
\end{equation}
where $K_{F}$ stands for the free fermionic action and $K_{B}$ is the
corresponding bosonized version. The term $I[j^{F}]$, with $j_{\mu }^{F}={%
\bar{\psi}}\gamma _{\mu }\psi ,$ represents a generic current interaction.
The bosonizing field $\lambda $ is a scalar field $\phi $ in $(1+1)D$, and a
vector field $A_{\mu }$ in $(2+1)D$. Accordingly, $\varepsilon \partial
\lambda $ has to be read as $\varepsilon _{\mu \nu }\partial _{\nu }\phi $
or $\varepsilon _{\mu \nu \rho }\partial _{\nu }A_{\rho }$, respectively. It
is worth mentioning here that the mapping $\left( \ref{univ}\right) $
provides a unifying framework to derive universal transport properties of
both one and two-dimensional interacting fermionic systems \cite{b111}.

Similarly to the $(1+1)D$ case, where fermions can be associated to soliton
configurations in the dual massive sine-Gordon theory \cite{b2}, one would
like to understand the elementary fermionic modes in $(2+1)D$ in terms of
topological excitations in the bosonized dual theory. The latter is a gauge
theory whose quadratic part is given by a nonlocal Maxwell-Chern-Simons
(MCS) term \cite{b,b11,b1}. In particular, when a large mass expansion is
performed, the dominant term reduces to the usual local MCS action, namely

\begin{equation}
S(A)=\int d^{3}x\left( \frac{1}{2m}\mathcal{F}_{\mu }^{2}\;+\;\frac{i}{2\eta 
}A_{\mu }\mathcal{F}^{\mu }\;\right) \;  \label{mcs-act}
\end{equation}
where $m$ is proportional to the fermion mass and $\eta $ is the
Chern-Simons coefficient in the fermionic effective action.

In $(2+1)D$, it is a common wisdom to believe that fermions should be
related to vortices in the dual theory. The aim of this letter is to pursue
this investigation. Combining 't Hooft approach \cite{th} to the
quantization of extended objects in euclidean space-time with the
Hennaux-Teitelbolm work \cite{ht} on instantons in MCS theory, we shall be
able to show that vortices may appear as excitations with definite mass and
spin in a generalized MCS model. The relationship among vortices in MCS and
fermionic excitations will be analysed through Polyakov's spin action for
Bose-Fermi transmutation in $(2+1)D$ \cite{p}. 't Hooft framework is
particularly adapted whenever the Mandelstam operators are not known. As an
example,  it has been successfully used to obtain a covariant quantization
for the soliton excitations of the Skyrme model \cite{ma}. We also point out
that the finite action requirement for vortex configurations is fulfilled by
introducing a suitable nonlocal Maxwell term.

The present letter is organized as follows. In Sect.2 we study MCS vortex
solutions in the presence of an instanton anti-instanton background. Sect.3
is devoted to the vortex quantization through the corresponding correlation
functions and to the analysis of Polyakov's term. In Sect.4, the nonlocal
MCS case is discussed.

\section{Vortices in Maxwell Chern-Simons}

In recent works \cite{d,kovner} the existence of vortex solutions in
Maxwell-Chern-Simons (MCS) in the presence of singularities has been
discussed. These singularities turn out to be related to the continuum limit
of a compact lattice version of the theory. The resulting classical solution
to the equations of motion displays the behavior of a vortex. Although this
configuration could be interpreted as a kind of energy lump due to its fast
decay given by the MCS topological mass, the corresponding total energy has
a mild logarithmic divergence in the ultraviolet region \cite{kovner}. In
addition, the vortex is pinned around the position of the singularity, which
is introduced as an external fixed source. In order to promote this field
configuration to a particle-like excitation we have to give translational
degrees of freedom to the vortex and render its energy finite. Also, the
vortex propagator should be well behaved, without unphysical modes.

Following 't Hooft procedure, the vortex propagation in euclidean space is
obtained by integrating over configurations where a vortex excitation is
created out of the vacuum at a space-time point $x_{1\mathrm{\ }}$ and after
an intermediate propagation is annihilated at $x_{2}$. Before $x_{1}$ and
after $x_{2}$ the topological charge vanishes, while it is nonvanishing in
between due to the existence of the vortex. Therefore, suitable instanton
anti-instanton singularities have to be introduced at $x_{1}$ and $x_{2}$ in
order to match these inequivalent topological configurations. In the present
three-dimensional case these singularities can be seen as a monopole
anti-monopole pair \cite{ht,pi} for the dual field strength configuration $%
\mathcal{F}_{\mu }=(1/2)\varepsilon _{\mu \nu \rho }F^{\nu \rho }$, located
at $x_{1}$ and $x_{2}\,$, respectively. One possible action describing the
coupling of this pair with the MCS field is given by

\begin{equation}
S(A,J)=\int d^{3}x\left( \frac{1}{2m}(\mathcal{F}_{\mu }+J_{\mu })^{2}\;+\;%
\frac{i}{2\eta }A_{\mu }\mathcal{F}^{\mu }\;\right) \;,  \label{in-act}
\end{equation}
with 
\begin{equation}
J^{\mu }(x)=\int_{\gamma }dy^{\mu }\delta ^{3}(x-y)\;,  \label{j}
\end{equation}
where $\gamma $ is an open smooth string running from $x_{1}$ to $x_{2}$

\begin{equation}
\partial ^\mu J_\mu =\delta ^3(x-x_1)-\delta ^3(x-x_2)\;.  \label{dj}
\end{equation}
The equations of motion are easily worked out and yield \cite{d}

\begin{eqnarray}
\mathcal{F}_{\mu }^{\mathrm{cl}} &=&-J_{\mu }+\mathcal{R}_{\mu }\mathcal{\;}
\label{sol} \\
\mathcal{R}_{\mu } &=&\frac{1}{4\pi }\left( \frac{m^{2}}{\eta ^{2}}\delta
_{\mu \alpha }-i\frac{m}{\eta }\varepsilon _{\mu \alpha \beta }\partial
^{\beta }\right) \int_{\gamma }dy^{\alpha }\,\frac{e^{-\frac{m}{\eta }\left|
x-y\right| }}{\left| x-y\right| }\;.  \nonumber
\end{eqnarray}
The term $\mathcal{R}_{\mu }$ in the above expression represents a vortex
configuration propagating from $x_{1}$ to $x_{2},$ having both magnetic and
electric field. We observe that, due to the presence of the exponential
factor in eq.$\left( \ref{sol}\right) $, $\mathcal{R}_{\mu }$ is localized
around the curve $\gamma $, on a scale of the order of $1/m.$ We also note
that the Bianchi identity $\partial ^{\mu }\mathcal{F}_{\mu }^{\mathrm{cl}%
}=0 $ implies that $\partial ^{\mu }\mathcal{R}_{\mu }=\delta
^{3}(x-x_{1})-\delta ^{3}(x-x_{2}).$ Therefore, the flux $\Phi $ of the
nonsingular part $\mathcal{R}^{z}$ of the magnetic field, computed through
any constant time plane $\Sigma $ located between $x_{1}$ and $x_{2}$, is 
\begin{equation}
\Phi =\int_{\Sigma }d^{2}x\;\mathcal{R}^{z}=\oint dS_{\mu }\mathcal{R}^{\mu
}=1\;,  \label{flux}
\end{equation}
where the second equality follows by closing $\Sigma $ with the addition of
a surface at infinity giving no contribution due to the exponential decay of 
$\mathcal{R}_{\mu }$.

The static limit corresponds to a configuration where the vortex is created
in the far past and annihilated in the far future, and it always sits at the
same position, that is, the associated string $\gamma $ is an infinite
straight line along the euclidean time-axis, identified with the $z-$axis.
In this case, eq.(\ref{sol}) reproduces the vortex profile discussed in ref.%
\cite{kovner}. In particular, for the magnetic field we get

\begin{equation}
\mathcal{F}_{z}^{\mathrm{cl}}=-\delta ^{(2)}(x)+\frac{1}{2\pi }K_{0}(\frac{m%
}{\eta }\rho )\;,  \label{b-field}
\end{equation}
with $K_{0}$ being the Bessel function and $\rho $ the radial coordinate in
the $(x,y)-$ plane. Also, the point-like singularity introduced in \cite
{kovner}, where the vortex is pinned, is nothing but the intersection of the
string with the constant time plane $\Sigma $.

\section{Quantization of the MCS vortices}

Following 't Hooft prescription \cite{th}, in order to compute the vortex
propagator we have to path integrate over all physical inequivalent
configurations representing the creation, propagation  and annihilation of
the vortex. Therefore, we integrate over the gauge fields and all possible
strings, and define the two-point vortex correlation function as 
\begin{equation}
\mathcal{G}(x_{1}-x_{2})=\int D\gamma \int DA\;e^{-S(A,J)}=\int D\gamma
\;e^{-\Gamma _{\gamma }\;},\;  \label{corr}
\end{equation}
where $\Gamma _{\gamma }$ represents the effective action obtained by
integrating over all gauge configurations in a fixed string background. The
presence of the measure $D\gamma $ is natural in a path integral approach 
\cite{p}, being in fact needed in order to ensure the string independence of 
$\mathcal{G}(x_{1}-x_{2}).$ This prescription should guaranty the locality
of the quantum vortex field operators whose expectation value has to be
identified with $\mathcal{G}(x_{1}-x_{2})$, although, in general, a closed
form for these operators is not known.

In the pure Maxwell case, corresponding to the limit $m\rightarrow 0$, $%
\Gamma _{\gamma }$ turns out to be independent from the particular Dirac
string joining the singularities \cite{th}

\begin{equation}
\Gamma _{\gamma }^{\mathrm{Max}}\propto \frac{1}{\left| x_{1}-x_{2}\right| }%
\;,  \label{max-eff}
\end{equation}
meaning that here the string is not observable. The integration over the
paths is now trivial and results in a pure normalization factor. The
path-independence of $\Gamma _{\gamma }^{\mathrm{Max}}$ allows us to deform
the original $\gamma $ into two strings $\gamma _{1},$ $\gamma _{2}$, where $%
\gamma _{1}$ goes from $x_{1}$ to $\infty $ and $\gamma _{2}$ from $\infty $
to $x_{2}$. In this case, the vortex correlation function in eq.$\left( \ref
{corr}\right) $ can be written in terms of Mandelstam variables $\mu (\gamma
_{1})$ , $\overline{\mu }(\gamma _{2})$, according to

\begin{eqnarray}
\mathcal{G}^{\mathrm{Max}}(x_{1}-x_{2}) &=&\mathcal{N}\int DA\;\mu (\gamma
_{1})\,\overline{\mu }(\gamma _{2})\,e^{-\frac{1}{2m}\int d^{3}x\,\mathcal{F}%
^{2}}\;,  \nonumber \\
\mu (\gamma _{1}) &=&e^{-\frac{1}{m}\int_{\gamma _{1}}dx^{\mu }\mathcal{F}%
_{\mu }\;}\;\;,\;\;\overline{\mu }(\gamma _{2})=e^{-\frac{1}{m}\int_{\gamma
_{2}}dx^{\mu }\mathcal{F}_{\mu }\;}.  \label{mand}
\end{eqnarray}
The string independence of the effective action $\left( \ref{max-eff}\right) 
$ corresponds to the well established locality properties of the Mandelstam
operators, in models containing pure Maxwell terms \cite{m}. Coming back to
the MCS case, it is easy to convince oneself that the effective action $%
\Gamma _{\gamma }\;$ in eq.$\left( \ref{corr}\right) $ has a nontrivial
dependence on $\gamma .$ Therefore, as the string is now observable, we have
to integrate over all paths, according to the general definition $\left( \ref
{corr}\right) .$ On physical grounds, this amounts to take into account all
possible intermediate processes representing the vortex propagation. We
underline that in this case an explicit expression for the vortex operators
is not available. However, the knowledge of the vortex propagator is
sufficient to characterize the physical properties of the vortex at the
quantum level.

As the integration over the gauge fields in eq.$\left( \ref{corr}\right) $
is quadratic, we obtain 
\begin{equation}
\Gamma _{\gamma }=S(A^{\mathrm{cl}},J)\;  \label{geffec}
\end{equation}
where $A^{\mathrm{cl}}$ is a vector potential for the saddle point
configuration $\mathcal{F}^{\mathrm{cl}}$ in eq.$\left( \ref{sol}\right) .$
After performing the space-time integral, $\Gamma _{\gamma }$ can be cast in
the form of a double-line integral over the curve $\gamma $, with a kernel
which is found to be localized on a scale of the order of $1/m$ (see eq.$%
\left( \ref{jj}\right) $ in Sect.4). For well separated $x_{1}$ and $x_{2}$,
and smooth strings, the effective action $\Gamma _{\gamma }$, up to order $%
1/m$, is 

\begin{equation}
\Gamma _{\gamma }\sim \lambda mL+\frac{\mathrm{const}}{m}\int_{0}^{L}ds\,%
\frac{de^{\alpha }(s)}{ds}\frac{de^{\alpha }(s)}{ds}\;,  \label{improv}
\end{equation}
where $L$ is the length of the curve $\gamma ,$ $e^{\alpha }(s)$ is the
tangent vector $dy^{\alpha }/ds$ and the parameter $s$ is defined through
the relation $e^{\alpha }(s)e^{\alpha }(s)=1$. The factor $\lambda $ is 
logarithmic divergent \cite{kovner}, and will be discussed in the next
section.

Notice that the presence of the second term in $\left( \ref{improv}\right) $
is in fact already known \cite{p} and takes into account velocity
correlations at different points along $\gamma $. In order to obtain the
vortex propagator $\mathcal{G}(x_{1}-x_{2})$ it remains to perform the
integration over all possible paths $\gamma $ with fixed end-points. This
integration can be found in \cite{p}, yielding as final result the
Klein-Gordon propagator.

The spinless character of this excitation is due to the complete
cancellation of all imaginary terms of the kind 
\begin{equation}
S_{\gamma }=\frac{1}{4\pi }\int_{\gamma }dx^{\alpha }\int_{\gamma }dy^{\beta
}\varepsilon _{\mu \alpha \beta }\,\partial _{\mu }^{x}\frac{1}{|x-y|}\;,
\label{sl}
\end{equation}
arising from the presence of the Chern-Simons action. Observe that, for
closed $\gamma $, this expression is known as the self-linking of the curve.

It is worth underlining that, depending on the coupling between the string
and the MCS gauge potential, different kinds of correlation functions will
be obtained, leading to different quantum numbers for the corresponding
vortex excitations. For instance, if instead of $\left( \ref{in-act}\right) $
one considers the more general coupling 
\begin{equation}
S(A,J)=\int d^{3}x\left( \frac{1}{2m}(\mathcal{F}_{\mu }+J_{\mu })^{2}\;+\;%
\frac{i}{2\eta }A_{\mu }\mathcal{F}^{\mu }+i\vartheta A_{\mu }J^{\mu
}\;\right) \;,  \label{sec-act}
\end{equation}
for the leading terms of the effective action $\Gamma _{\gamma }$ one gets 
\begin{equation}
\Gamma _{\gamma }\sim \lambda mL+\frac{i}{2}\eta \vartheta ^{2}S_{\gamma }\;.
\label{n-r}
\end{equation}
In particular, for $\eta \vartheta ^{2}=2\pi ,$ Polyakov's Bose-Fermi
transmutation occurs and the vortex propagator turns out to be that of a
spin one-half fermionic excitation \cite{p,ha} 
\begin{equation}
\int d^{3}p\frac{1}{\sigma ^{\mu }p_{\mu }+\lambda m}e^{ip(x_{1}-x_{2})}
\label{dirac}
\end{equation}
where $\sigma _{\mu }$ are the Pauli matrices. With respect to the spinor
index structure of this propagator we refer the reader to the original work 
\cite{p}. In this regard, it is useful to point out that the functional
integration in eq.$\left( \ref{corr}\right) $ should be equipped with
appropriate fixed boundary conditions around the monopole anti-monopole
singularities, carrying a representation of the rotation group. At the
locations of these singularities vortices with given quantum numbers will be
created and destroyed. This will lead to the correct index structure for the
final expression of the propagator. This framework has been worked out in
ref.\cite{ma} in the case of skyrmions.

\section{Vortices in nonlocal MCS models}

So far, we have seen that vortex configurations are present in MCS theory
when a nontrivial instanton anti-instanton background is introduced.
Depending on the coupling with the string, the vortex quantum numbers may
correspond to a bosonic or a fermionic excitation. However, as it has been
already pointed out in \cite{kovner}, the energy of this configuration
displays an ultraviolet logarithmic divergence. The aim of this section is
to face this problem. One possibility in order to have a finite action
configuration is that of introducing nonlocal terms in the action, whose
effect is that of properly regularizing the ultraviolet region. For
instance, this can be done by modifying the Maxwell term in $\left( \ref
{sec-act}\right) $ according to 
\begin{equation}
S(A,J)=\int d^{3}x\left( \frac{1}{2}(\mathcal{F}_{\mu }+J_{\mu })\hat{O}(%
\mathcal{F}^{\mu }+J^{\mu })\;+\;\frac{i}{2\eta }A_{\mu }\mathcal{F}^{\mu
}+i\vartheta A_{\mu }J^{\mu }\right) \;,  \label{nl-act}
\end{equation}
where $\hat{O}$ is a nonlocal operator associated with a kernel $O(x-y)$ 
\[
\left[ \hat{O}\mathcal{F}\right] (x)=\int d^{3}yO(x-y)\mathcal{F}(y)\;.
\]
We also require that the Fourier transform 
\begin{equation}
\widetilde{O}(k)=\int d^{3}x\,e^{-ikx}O(x)  \label{f}
\end{equation}
is positive definite.

The local Maxwell term is recovered by taking $O(x-y)=(1/m)\,\delta
^{(3)}(x-y)$. We remark here that nonlocal MCS models appear in a natural
way in the context of bosonization \cite{b11}. Indeed, these terms arise
from the evaluation of the massive fermionic determinant in a generic
background. We also observe that the presence of a current-current
interaction in the starting fermionic action will produce in the bosonized
action an additional nonlocal Maxwell term, which follows from the universal
bosonization rule $\left( \ref{univ}\right) $, namely 
\begin{equation}
\frac{1}{2}\int d^{3}x\,d^{3}y\,j_{\mu }^{F}(x)G(x-y)j_{\mu
}^{F}(y)\leftrightarrow \frac{1}{2}\int d^{3}x\,d^{3}y\,\mathcal{F}_{\mu
}(x)G(x-y)\mathcal{F}^{\mu }(y)\;.  \label{nluniv}
\end{equation}

Coming back to the nonlocal MCS action $\left( \ref{nl-act}\right) $, the
corresponding classical vortex profile gets modified according to 
\begin{equation}
\mathcal{F}_\mu ^{\mathrm{cl}}=-J_\mu +\frac{(1-\eta \vartheta )}{1-\eta ^2%
\hat{O}^2\partial ^2}\left( J_\mu +i\eta \hat{O}\varepsilon _{\mu \nu \rho
}\partial _\nu J_\rho \right) \;.  \label{nl-sol}
\end{equation}
Upon substitution of this expression in eq.$\left( \ref{nl-act}\right) $ one
obtains

\begin{eqnarray}
S(A^{\mathrm{cl}},J) &=&\frac{i}{2}\eta \vartheta ^{2}S_{\gamma }+\frac{1}{2}%
(1-\vartheta \eta )^{2}\int d^{3}xJ_{\mu }\frac{\hat{O}}{1-\eta ^{2}\hat{O}%
^{2}\partial ^{2}}J_{\mu }  \nonumber \\
&&+\frac{i}{2\eta }(1-\vartheta \eta )^{2}\int d^{3}xJ_{\mu }\frac{\hat{O}%
^{2}}{1-\eta ^{2}\hat{O}^{2}\partial ^{2}}\varepsilon _{\mu \nu \rho
}\partial _{\nu }J_{\rho }  \label{jj}
\end{eqnarray}
We note that the real part of the action is positive. Also, in the static
limit in which $\gamma $ is an infinite straight line coinciding with the $z-
$axis, the action per unit length turns out to be 
\begin{equation}
\frac{1}{2}(1-\vartheta \eta )^{2}\int \frac{d^{2}k}{(2\pi )^{2}}\,\frac{%
\mathbf{\widetilde{O}}}{1+\eta ^{2}\mathbf{\widetilde{O}}^{2}\mathbf{k}^{2}}%
\;,
\end{equation}
where the quantities in boldface correspond to the two-dimensional
projection $k\rightarrow (\mathbf{k,}0)$. In the local case ($\widetilde{O}%
=1/m$) this expression contains a mild logarithmic ultraviolet divergence 
\cite{kovner}. However, in the case where $\widetilde{O}$ behaves in the uv
region as $k^{\alpha }\;(\alpha >0),$ the action per unit length is rendered
finite, no matter how small $\alpha $ is.

\section{Conclusions}

Following 't Hooft procedure, we have studied vortex correlation functions
in MCS models considering different couplings between the gauge fields and
the string associated with the instanton anti-instanton pair. This string
arises in the continuum limit of a compact lattice version of the theory 
\cite{kovner,d}.

With the exception of the pure Maxwell type case, the string is observable.
Therefore, we have defined vortex correlation functions by path integrating
over both the gauge fields and the string. This corresponds to take into
account the vortex translational degrees of freedom. It is the integration
over the string which finally leads to a well behaved propagator, without
unphysical poles.

Concerning the bosonization of $(2+1)D$ fermionic systems we remind that,
for large $m$, the dominant term in the bosonized action corresponds to the
local MCS \cite{b}. Furthermore, we have been able to see that the coupling
in eq.$\left( \ref{sec-act}\right) $ leads to a vortex excitation with spin $%
1/2$, whenever the condition $\eta \vartheta ^{2}=2\pi $ is satisfied.
Although a direct derivation of the bosonization formula for fermion
propagators has not yet been obtained, this result gives a strong indication
that the elementary fermionic excitations correspond indeed to vortices in
the dual theory.

These vortex configurations have been generalized to the case in which a
nonlocal Maxwell term is present. We have shown that this kind of term could
improve the ultraviolet behavior so as to render the vortex energy finite.

On the other hand, for $\eta \vartheta ^{2}=2\pi $, the possibility of
identifying vortex and fermionic correlation functions together with the
universal bosonization rule $\left( \ref{nluniv}\right) $ could give a
useful framework to analyse the spectrum of the excitations for interacting
fermionic systems. While in the local MCS case the localization of the
vortex on a scale of the order $1/m$ leads to the existence of a pole in the
vortex propagator due to eq.$\left( \ref{n-r}\right) $, in the nonlocal
case, depending on the fermionic interaction kernel $G(x-y)$ in eq.$\left( 
\ref{nluniv}\right) $, the vortex profile $\left( \ref{nl-sol}\right) $
could spread out. This would imply the breaking of the validity of the long
distance approximation $\left( \ref{n-r}\right) $. This may result in the
absence of the pole in the propagator, meaning that the quasiparticle
picture could be destabilized by the interaction among fermions.

\section{Acknowledgements}

The Conselho Nacional de Desenvolvimento Cient\'{\i }fico e Tecnol\'{o}gico
CNPq-Brazil, the Funda{\c{c}}{\~{a}}o de Amparo {\`{a}} Pesquisa do Estado
do Rio de Janeiro (Faperj) and the SR2-UERJ are acknowledged for the
financial support. \newline
\newline

\end{document}